\documentclass[prl,floatfix,twocolumn,superscriptaddress,showpacs,preprintnumbers,longbibliography,nofootinbib]{revtex4-2}
\usepackage{color}
\usepackage[utf8]{inputenc}
\usepackage[usenames,dvipsnames]{xcolor}
\usepackage{amsmath}
\usepackage{amssymb}
\usepackage{graphicx}
\graphicspath{{graphics/}}
\usepackage{epsfig}
\usepackage{dcolumn}
\usepackage{bm}
\usepackage{mathrsfs}
\usepackage{multirow}
\usepackage[all]{xy}
\usepackage{pbox}
\usepackage{lipsum}
\usepackage{nicefrac}
\usepackage{verbatim}
\usepackage{braket}
\usepackage{dsfont}
\usepackage{array}
\usepackage{makecell}
\usepackage{tabularx}
\usepackage{isotope}
\usepackage{xr}
\usepackage{amsthm} 
\usepackage[normalem]{ulem} 
\usepackage{physics}
\usepackage{float}
\usepackage{placeins}
\usepackage{svg}

\usepackage[colorlinks=true,linkcolor=blue,urlcolor=blue,citecolor=blue,pdfusetitle]{hyperref}

\usepackage{tikz}

\begin{document}
\title{Getting large-scale quantum neural networks ready for quantum hardware}

\author{Mario Boneberg}
\affiliation{Institut f\"ur Theoretische Physik, Universit\"at Tübingen and Center for Integrated Quantum Science and Technology, Auf der Morgenstelle 14, 72076 T\"ubingen, Germany}
\author{Simon Kochsiek}
\affiliation{Institut f\"ur Theoretische Physik, Universit\"at Tübingen and Center for Integrated Quantum Science and Technology, Auf der Morgenstelle 14, 72076 T\"ubingen, Germany}
\author{Igor Lesanovsky}
\affiliation{Institut f\"ur Theoretische Physik, Universit\"at Tübingen and Center for Integrated Quantum Science and Technology, Auf der Morgenstelle 14, 72076 T\"ubingen, Germany}
\affiliation{School of Physics and Astronomy and Centre for the Mathematics and Theoretical Physics of Quantum Non-Equilibrium Systems, The University of Nottingham, Nottingham, NG7 2RD, United Kingdom}

\begin{abstract}
Quantum neural networks generalize classical artificial neural networks into the quantum domain. They are formulated as parameterized quantum circuits which are optimized by measuring and minimizing a suitably chosen loss function. The core challenge in understanding, implementing and ultimately using quantum neural networks is that they represent many-body systems with an exponentially large Hilbert space, in combination with a large parameter search space. Moreover, noise --- which is inherent to any quantum measurement --- sets practical limits for the estimation of training loss. Here, we study physics-informed large-scale quantum neural networks that are trained through a finite number of noisy loss function measurements. We show that this architecture permits the construction of nontrivial decision boundaries that enable the classification of quantum states through measuring an order parameter. Our approach can directly process quantum data that is output from quantum simulators and computers and is well suited for implementation on current hardware. Moreover, owed to a close link between the neural network dynamics and the evolution of Markovian open many-body quantum systems, one may expect a certain robustness to noise, which is ubiquitous in the current NISQ era.
\end{abstract}
\maketitle
\noindent {\bf Introduction.---}
\begin{figure}[t]
    \centering
    \includegraphics[width=\linewidth]{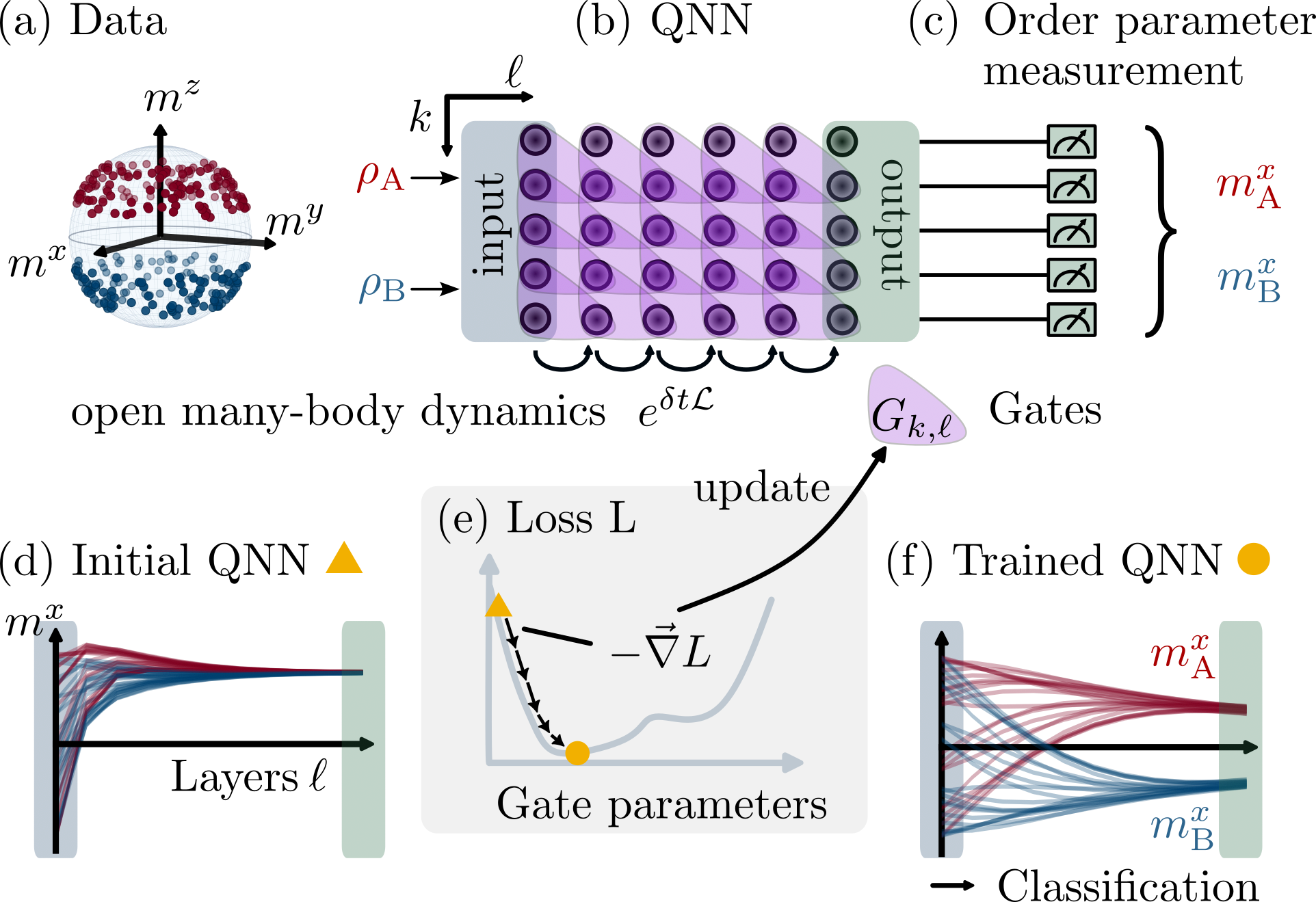}
    \caption{\textbf{Physics-informed quantum neural network architecture and training.} (a) Structured dataset that is used to train the network. It consists of labeled quantum states. In the example shown they are represented on the Bloch sphere, spanned by the magnetization components $m^{x,y,z}$ [see main text] and fall into two classes: A (red dots, northern hemisphere) and B (blue dots, southern hemisphere). (b) The corresponding states ($\rho_A$ and $\rho_B$) are fed into the first layer of qubits of a quantum neural network. Local, translation-invariant gates, applied between consecutive layers, propagate these states layer-by-layer to the rightmost (output) layer according to an open quantum many-body dynamics generated by a Lindbladian $\mathcal{L}$. (c) The magnetization $m^x$ serves as an order parameter and we measure it on the output layer for the two kinds of input states which yields $m_{\mathrm{A}}^x$ and $m_{\mathrm{B}}^x$. The order parameter is used to compute the loss function for training the network and ultimately allows to classify the input. (d) In the untrained network measuring the output magnetization in x-direction does not allow to infer the label of the input states. (e) The loss function quantifies for different parameterizations of the gates how far the network is from classifying the data. Thus, the output values for the order parameter in the untrained network determine a specific point on the loss landscape. The gradient of the loss at this point provides a direction in which we may update the gate parameters in order to move towards an improved network. (f) Iterating this update, the final (trained) network allows to classify input states by measuring the output magnetization in x-direction.}
    \label{fig:1}
\end{figure}
Quantum machine learning \cite{biamonte2017,schuld2021,schuld2015,schuld2022,cerezo2022,rebentrost2014,schuld2014} aims to combine the empirical success of classical machine learning algorithms, such as artificial neural networks \cite{bishop2006,nielsen2015,goodfellow2016,lecun2015}, with the potential computational advantages provided by quantum information processing \cite{nielsen2010,shor1997,grover1996}. A paradigmatic example are quantum neural networks (QNNs) \cite{killoran2019, mangini2021, cerezo2021a,beer2020} which employ parametrized quantum circuits whose gates are optimized with respect to a given task. The optimization part is performed on a classical computer using measurement outcomes from the circuit. This framework enables the identification of patterns in data consisting of quantum states which may, e.g., be the output of a quantum simulation, a quantum sensor or a quantum many-body experiment. These states live in a Hilbert space which is exponentially large in the number of their elementary constituents. Therefore, characterization of the full state via tomography is infeasible and QNNs necessarily operate in a regime which is intractable by classical approaches \cite{feynman1982,lloyd1996}. \\
QNNs in principle offer some resilience against decoherence \cite{cerezo2021a,sharma2020} and therefore may be already implementable on current noisy intermediate-scale quantum (NISQ) devices \cite{preskill2018,wecker2015,mcclean2016}. The reason for this robustness is that the classical variational optimization of QNNs, which entails the minimization of a loss function, can as well be carried out with noisy quantum circuits. This has been shown in recent experiments where QNN algorithms are executed successfully on quantum processors with small numbers of qubits \cite{pan2023,omalley2016,buonaiuto2024}. For large-scale QNNs, however, there are fundamental challenges that render the optimization procedure impractical. In particular so-called barren plateaus \cite{mcclean2018,anderson1967,ragone2024,larocca2025,cerezo2025,sharma2022,cerezo2021b}, i.e., loss functions that become exponentially flat, make it increasingly harder to compute optimization gradients. Computing an optimization step thus requires an exponential number of shots to suppress sampling noise \cite{scriva2024,kreplin2024,kim2025,kaminishi2026,recio-armengol2025}, which is infeasible. \\
In this work, we demonstrate that physics-informed large-scale QNNs \cite{gillman2022b,boneberg2023,boneberg2025,boneberg2025b} can be trained using loss functions which can be efficiently estimated through finite-shot measurements in order to classify quantum input data. The networks we consider for this purpose are layered quantum many-body systems [see Fig.~\ref{fig:1} for an illustration of the principle]. The first layer contains the input state, which is propagated forward layerwise by means of sequences of local, translation-invariant gates and generates an output state on the last layer. Despite the fully unitary construction, this realizes a discrete-time open quantum many-body dynamics on the level of the individual layers \cite{beer2020,bondarenko2020,beer2022,gillman2022b,boneberg2023}. Such dynamics may feature emergent phenomena, e.g. phase transitions and ergodicity breaking, and hence allows to classify quantum states \cite{gillman2022b,boneberg2025, boneberg2023,boneberg2025b}. Indeed, employing an order-parameter-based loss function we simulate large QNNs with tensor networks \cite{orus2014,paeckel2019,gillman2021b,boneberg2025} and show that they can be trained to classify synthetically generated quantum data. This demonstrates that our approach is capable of learning complex decision boundaries in data on NISQ hardware. \\

\noindent {\bf Physics-informed quantum neural networks.---}
The QNNs we consider here \cite{beer2020,bondarenko2020,lewenstein2021,beer2022,gillman2022b,boneberg2023,locher2023,sutter2025,boneberg2025} are defined on a rectangular lattice with $L+1$ vertical layers each containing $N$ sites [cf. Fig.~\ref{fig:1}(b)]. To each site a qubit with basis states $\ket{0}$ (vacuum) and $\ket{1}$ is associated. Layers are indexed by $\ell$ and the leftmost layer, $\ell = 0$, initially contains the input state $\rho_{\mathrm{in}}$, while all other layers are in the vacuum state. Vertical sites are labeled by $k$, starting from the top, $k=1$, and we apply local gates $G_{k,\ell}$, acting between layers $\ell -1, \ell$ and in a neighborhood of the $k$th site [cf. Fig.~\ref{fig:1}(b)]. This yields the output state 
\begin{equation}
    \rho_{\mathrm{out}} = \Tr_{\neq L} \left( \prod_{\ell=L}^1 \prod_{k=1}^N  G_{k,\ell}  \rho_{\mathrm{in}} \otimes \outerproduct{\mathbf{0}}_{\neq 0} \prod_{\ell=1}^L \prod_{k=N}^1  G_{k,\ell}^\dagger  \right)
\end{equation}
on the rightmost layer, where $\Tr_{\neq L}$ denotes the trace over all other layers and $\outerproduct{\mathbf{0}}_{\neq 0}$ the vacuum state on all but the first layer. \\
We link the computation of the output state to an open quantum many-body dynamics by parametrizing the local gates as \cite{gillman2022b,boneberg2023}
\begin{equation} \label{localgate}
    G_{k,\ell} = \mathrm{SWAP}_{k,\ell} \times e^{-i \sqrt{\delta t} \big( J_{k,\ell-1}  \sigma_{k,\ell}^+ + \text{h.c.}\big)} \times e^{-i \delta t H_{k,\ell - 1} },
\end{equation}
where $H_{k, \ell - 1}$ and $J_{k, \ell -1}$ are local hermitian operators acting in a neighborhood of site $k$ in layer $\ell -1$ and $\sigma_{k,\ell}^+ = (\sigma_{k,\ell}^x + i \sigma_{k,\ell}^y ) / 2$ is defined in terms of the Pauli matrices $\sigma_{k, \ell}^{\alpha}$, with $\alpha=I,x,y,z$, on site $k$ of layer $\ell$. The rightmost term then generates an evolution on layer $\ell - 1$ under $H_{k, \ell - 1}$ and the middle term evolves according to $J_{k, \ell -1}$. At the end, the operator $\mathrm{SWAP}_{k,\ell}$ swaps the state on site $k$ from layer $\ell -1$ to layer $\ell$ and generates a forward propagation. The hyperparameter $\delta t$ can be interpreted as a time step. A convenient way of expressing the dynamics -- which makes evident the connection to many-body physics -- is to consider the limit $\delta t \ll 1$ for which the input state $\rho_{\mathrm{in}}= \rho_0$ is evolved to the output state $\rho_{\mathrm{out}}= \rho_L$ layer-by-layer according to the recurrence relation \cite{gillman2022b,boneberg2023,boneberg2025,lorenzo2017,ciccarello2017,ciccarello2022,cattaneo2021,cattaneo2022}
\begin{equation} \label{recurrence}
    \rho_\ell\approx \exp (\mathcal{L}_{\ell -1}\delta t)[\rho_{\ell-1}], 
\end{equation}
with the \textit{Lindblad generator} \cite{lindblad1976,gorini1976,breuer2002,rivas2012}
\begin{align}\label{lindbladian}
     \mathcal{L}_{\ell -1}[\bullet] =& -i\left[\sum_{k=1}^N H_{k,\ell -1}, \bullet \right] \nonumber \\ 
     &+ \sum_{k=1}^N \left( J_{k,\ell-1} \bullet J_{k,\ell-1}^\dagger - \frac{1}{2} \left\{ J_{k,\ell-1}^{\dagger} J_{k,\ell-1}, \bullet \right\} \right) ,
\end{align}
the \textit{Hamiltonians} $H_{k,\ell -1}$ and the \textit{jump operators} $J_{k,\ell -1}$. In the following, we choose the latter two to be translation invariant nearest-neighbor operators with the forms
\begin{equation}\label{hamiltonians}
H_{k,\ell -1}=\sum_{\alpha_1,\alpha_2=I,x,y,z} h_{\alpha_1 \alpha_2} \sigma_{k-1,\ell - 1}^{\alpha_1}  \sigma_{k,\ell - 1}^{\alpha_2}
\end{equation}
and 
\begin{equation}\label{jumpoperators}
J_{k,\ell -1}=\sum_{\alpha_1,\alpha_2=I,x,y,z} j_{\alpha_1 \alpha_2} \sigma_{k-1,\ell - 1}^{\alpha_1}  \sigma_{k,\ell - 1}^{\alpha_2}.
\end{equation}
Here $h_{\alpha_1 \alpha_2}$ and $j_{\alpha_1 \alpha_2}$ are real and complex numbers, respectively, which we gather in the matrices $\mathbf{h}$ and $\mathbf{j}$.
Equation \eqref{lindbladian} represents a discrete version of an open quantum many-body dynamics which may feature emergent effects such as phase transitions, symmetry breaking and ergodicity breaking \cite{bartolo2016,minganti2023,lieu2020,buca2012,albert2014,sieberer2025,rose2016,ates2012,weimer2015,rotondo2018,boneberg2025,boneberg2025b}  that enables data classification. To characterize the output, we consider in the following the magnetization observables
\begin{equation}
    \hat{m}^\alpha = \frac{1}{2N} \sum_{k=1}^N \sigma_k^\alpha .
\end{equation} 
Evolving their expectation values $m^\alpha = \Tr ( \hat{m}^\alpha \rho )$ via Eq. \eqref{recurrence} through the network can contract distinct clusters of input states towards distinct output states \cite{boneberg2025,boneberg2025b} [cf. Fig.~\ref{fig:1}]. Thus, inputs may be classified by the magnetization on the output layer $m_{\mathrm{out}}^\alpha =  \Tr ( \hat{m}^\alpha \rho_{\mathrm{out} })$, which can be interpreted as an order parameter for the many-body system. Another practical advantage of magnetization observables is that they can be efficiently evaluated on a quantum device. Here, measuring the magnetization $\hat{m}^\alpha$ projects the output state onto an eigenstate of this measurement operator and yields an outcome according to probabilities determined by the Born's rule. Each such measurement outcome is called a single shot. Collecting $S$ shots, their mean, $m_{S, \mathrm{out}}^\alpha$, provides an estimate of the expectation value $m_{\mathrm{out}}^\alpha$. The central limit theorem applies and we therefore expect the scaling $\mathrm{std}_S(m^\alpha_\text{S,out}) \propto 1/\sqrt{SN}$ for the standard deviation around $m^\alpha_\text{out}$, where $N$ is the width of the network layers. We will exploit this for training our QNNs. 
\begin{figure*}[t]
    \centering
    \includegraphics[width=\textwidth]{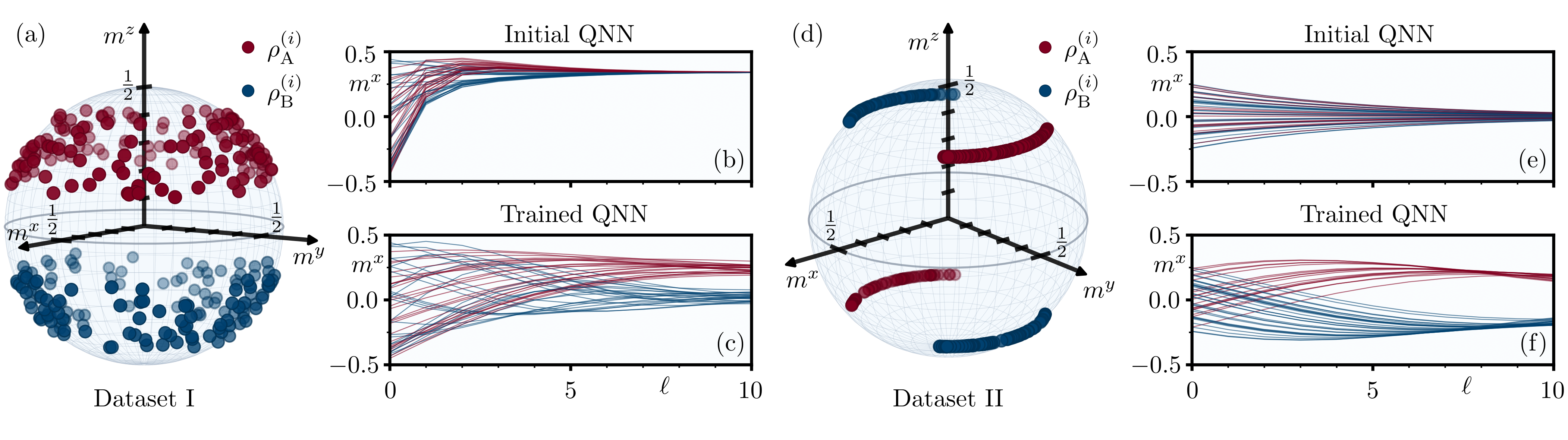} 
\caption{\textbf{Training on datasets I and II.} (a) Dataset I of quantum states represented on the Bloch sphere. Class A states (red dots) can be distinguished from class B states (blue dots) by measuring $\hat{m}^z$, i.e. populations. (b) Evolution of the expectation value $m^x$ through the layers of the initial QNN for the validation data. All trajectories converge towards a single stationary magnetization. (c) Evolution of $m^x$ for the trained QNN. Initial states with different labels have output magnetizations in different ranges, implying perfect classification on the validation set. The training algorithm thus found a solution which classifies the data in a basis where the original states were not distinguishable. (d) Dataset II. States are sampled from lines on the Bloch sphere and unlike the states of dataset I, they cannot be separated by a plane. Class A states (red dots) can only be distinguished from class B states (blue dots) by measuring both $\hat{m}^y$ and $\hat{m}^z$. (e) The initial QNN for the second dataset displays a unique output magnetization $m^x$ and hence does not classify inputs. (f) The trained network contracts states of different classes towards different output magnetization values and therefore allows to classify inputs based on measuring a single observable $\hat{m}^x$. For the simulations of the QNN evolution we employed matrix product operators with bond dimension $\chi_{\mathrm{MPO}}=16$ and matrix product states with bond dimension $\chi_{\mathrm{MPS}}=48$ \cite{SM, boneberg2025}. The training behavior of the QNN depends critically on the choice of hyperparameters and for all plots we have chosen a learning rate $\lambda=0.05$, a time step $\delta t = 0.1$ and a finite-difference step size $\epsilon =0.1$. For details on the function $f_r$ we refer to the supplemental material \cite{SM}.}
    \label{fig:2}
\end{figure*} \\
\noindent{\bf Quantum neural network training.---} To study how the classification capability of our QNNs is trained, we consider a setting where we have access to a dataset $\{\rho_{\nu_i}^{(i)}\}_{i=1}^{P}$ of $P$ quantum states, labeled by $\nu_i \in \{ \mathrm{A, B} \}$. These states are sampled randomly from an underlying, structured, distribution. They may, for instance, be the output of a quantum device or quantum experiment, which are directly fed into the input layer. For the purpose of this work --- in which we simulate quantum dynamics on a classical computer --- we choose them to be pure, translation-invariant product states, $\rho_{\nu_i}^{(i)} = \bigotimes_{k=1}^N \outerproduct*{\psi_{\nu_i}^{(i)}}{\psi_{\nu_i}^{(i)}}$, sampled along predefined directions $(m^x,m^y,m^z)^T$ on the Bloch sphere [cf.\ Fig.~\ref{fig:1}(a)]. These states are straightforward to prepare and their evolution can be simulated efficiently. The ultimate goal for the network is to encode the structure of the distribution such that it \textit{generalizes} from the available dataset and classifies arbitrary quantum states according to A and B. In order to achieve this, we employ a training algorithm which allows to find the optimal network gates in an automated way. We first describe the general scheme and later discuss two concrete examples. \\
To achieve classification on the output layer of the QNN, we measure only the magnetization $\hat{m}^x$, i.e. probe a single direction on the Bloch sphere. The network then needs to find the parameter matrices $\mathbf{h}, \mathbf{j}$ in Eqs.~\eqref{hamiltonians},\eqref{jumpoperators} such that it implements a nontrivial quantum dynamics which, for input state $\rho_{\mathrm{in}} = \rho_{\nu_i}^{(i)}$, yields a value $m_{S, \mathrm{out}}^{x,[i]} = a$ if the label is $\nu_i =\mathrm{A}$ and a value $m_{S, \mathrm{out}}^{x,[i]}=b$ if the label is $\nu_i= \mathrm{B}$. Training the network to fulfill this task corresponds to minimizing a loss function over the gate parameters $\mathbf{h}, \mathbf{j}$, which are initially chosen random [cf.\ Fig.~\ref{fig:1}(e)]. Particularly as we work with quantum states, it is favorable to use a loss function which can be computed with minimal knowledge. For this reason, we  employ the \textit{contrastive loss}
\allowdisplaybreaks
\begin{align}
    L(h, j) &= \frac{1}{P^2} \sum_{i,l=1}^{P} \bigg[  y_{il} \left( m_{S, \mathrm{out}}^{x,[i]} - m_{S, \mathrm{out}}^{x,[l]} \right)^2 \nonumber \\ 
    & +  (1 - y_{il} ) \max \left( 0, d - |m_{S, \mathrm{out}}^{x,[i]} - m_{S, \mathrm{out}}^{x,[l]}| \right)^2 \bigg] ,
\end{align}
where $d \in [0, 1 ]$, $y_{il}=1$ if $\nu_i=\nu_l$ (same class) and $y_{il}=0$ otherwise. This loss function uses only relative information between outputs: The first term penalizes differences in output magnetization for input states of the same class exclusively. This eventually leads to a contraction of outputs within a class during training. The second term penalizes output magnetizations for input states of different classes that are not resolved by at least a distance $d$. This leads to a separation of the two classes. We minimize $L$ using a stochastic gradient descent variant \cite{ruder2016}, where we compute the gradient of the loss function with respect to a randomly selected minibatch of the dataset via the finite-difference components
\begin{equation}
    \nabla_{\alpha_1, \alpha_2, \beta} L ( \mathbf{h}, \mathbf{j} ) \approx  \frac{ L \left( (\mathbf{h}, \mathbf{j} ) + \epsilon \mathbf{E}^\beta_{\alpha_1 \alpha_2} \right) - L( \mathbf{h}, \mathbf{j} ) }{\epsilon} .
\end{equation}
Here the parameter $\epsilon $ is small, $\beta =0 \ \mathrm{or} \ 1$, $\mathbf{E}_{\alpha_1 \alpha_2}^1 = (\mathbf{E}_{\alpha_1 \alpha_2},0)$, $\mathbf{E}_{\alpha_1 \alpha_2}^2 = (0, \mathbf{E}_{\alpha_1 \alpha_2})$ and $\mathbf{E}_{\alpha_1 \alpha_2}$ is the matrix with a one at position $(\alpha_1, \alpha_2)$ and zeros otherwise. The negative gradient gives us the direction of steepest descent of the loss function in parameter space [see Fig.~\ref{fig:1}(e)]. We then move into this direction by a small amount 
\begin{equation}
(\mathbf{h}, \mathbf{j}) \longrightarrow (\mathbf{h}, \mathbf{j}) - \lambda f_r(\nabla L(\mathbf{h}, \mathbf{j}))
\end{equation}
controlled by the \textit{learning rate} $\lambda$ and the vector-valued function $f_r$ (see supplemental material (SM) \cite{SM} for details). We repeat this procedure multiple times and in each iteration $r$ we sample a different minibatch. The function $f_r$ controls the update by keeping track of past gradients (momentum) and incorporating a lookahead \cite{ruder2016,kingma2014,dozat2016}. In classical machine learning, such variants of gradient descent have empirically proven to be highly efficient optimization strategies with strong performance \cite{bishop2006,lecun2015,nielsen2015,goodfellow2016}. In quantum machine learning, on the other hand, there are well-known fundamental issues which can prevent QNNs from finding meaningful optimization directions due to large flat areas, so-called barren plateaus \cite{mcclean2018,anderson1967,ragone2024,larocca2025,cerezo2025,sharma2022,cerezo2021b}, in the loss function landscape. As we demonstrate below, however, our architectures can in fact be trained successfully. \\

\noindent {\bf Training and Classification from few-shot measurements.---} 
Our QNNs are composed of $550$ qubits which are distributed over $N=50$ sites and $L+1=11$ layers. We simulate their evolution by employing tensor network techniques \cite{orus2014,paeckel2019,gillman2021b,boneberg2025} and conduct $5000$-shot measurements of $\hat{m}^x$ on the output layer (see SM \cite{SM} for details). We study two datasets that differ in the complexity of their decision boundaries. Since the data states are translation-invariant products of pure states, they are fully determined by a Bloch vector $\Vec{m}_i$ with $|\Vec{m}_i|=1/2$ through $\outerproduct*{\psi_{\nu_i}^{(i)}}{\psi_{\nu_i}^{(i)}} = \mathbb{I}/2 + m_i^x \sigma^x + m_i^y \sigma^y + m_i^z \sigma^z$. \\
\textit{Dataset I.} For the first dataset, as illustrated in Fig.~\ref{fig:2}(a), we randomly sampled $300$ states where $\rho_{\mathrm{A}}^{(i)}$ (red dots) are characterized by $m_i^z \in [0.15, 0.4]$ and $\rho_{\mathrm{B}}^{(i)}$ (blue dots) by $ m_i^z \in [-0.4, -0.15]$ (see \cite{SM} for the concrete sampling strategy). The two classes may therefore be distinguished by measuring $\hat{m}^z$, which probes classical probability distributions (populations) in the computational basis. For networks with $N=50$, this can be done efficiently as fluctuations around the mean $m_S^z=m_{5000}^z$ are small due to the central limit theorem. However, the classes cannot be distinguished by measuring $\hat{m}^x$, which probes quantum coherences --- this is the task we set for the network. We split the $300$ data states into a subset of $P=250$ states with which we train the network and a subset of $50$ validation states which are not seen by the algorithm during training and for which we will test the classification capability of the network by computing a so-called \textit{validation loss}. We set a distance $d=0.25$ and optimize $10$ parameters (see SM \cite{SM}). The parameters are initialized randomly, such that the network initially maps all validation states to a single stationary output magnetization \cite{schirmer2010,nigro2019} as can be seen in Fig.~\ref{fig:2}(b) [cf. Fig.~\ref{fig:1}(d)]. We then train the network for $50$ rounds with minibatches of size $25$. Both the training loss and the validation loss decrease and we select the parameters for which the validation loss is minimal (see SM \cite{SM} for the behavior of the loss functions). The resulting QNN maps states with different labels to distinct ranges of output magnetization $m_{\mathrm{out}}^{x}$ [see Fig.~\ref{fig:2}(c) and Fig.~\ref{fig:1}(f)], demonstrating that the network has learned to extract quantum features --- superpositions and phase information --- for classification.\\
\textit{Dataset II.} The second dataset, shown in Fig.~\ref{fig:2}(d), exhibits a more complex decision boundary. Class A states (red dots) are sampled along two lines on the Bloch sphere with $\mathrm{sign}(m^y m^z)=-1$ and the states belonging to class B (blue dots) along two lines with $\mathrm{sign}(m^y m^z)=1$ (see SM \cite{SM}). By construction, for each state on a line belonging to class A there exist two states on lines belonging to class B: one with the same $m^x$ and $m^y$ values and one with the same $m^x$ and $m^z$. The classes can therefore not be separated by a single linear cut on the Bloch sphere and the class of a randomly sampled state must be determined by measuring both, $\hat{m}^y$ and $\hat{m}^z$. The task for the network is thus to classify the states by a single observable $\hat{m}^x$.
Starting from a random initialization (see SM \cite{SM}) that does not classify inputs, we set a distance $d=0.35$ in the loss function and train $11$ parameters with minibatches of size 23 for $20$ rounds. The network with the minimal validation loss achieves classification [see Fig.~\ref{fig:2}(f)] and hence found a representation which reduces the complexity from measuring two observables to measuring a single observable for distinguishing the labels.  \\

\noindent {\bf Summary and outlook.---} 
We have shown that QNNs can be trained on large scale to achieve classification of quantum input data. Using tensor network simulations of physics-informed QNNs containing $550$ qubits, we demonstrated that a randomly initialized network learns binary classification tasks from $5000$ measurement shots per output state, on synthetically generated datasets.
Central to our approach is a contrastive loss function, which is computed based on an order parameter observable that is expected to obey a central limit theorem, enabling efficient estimation and training on large-scale. Since our QNNs effectively implement a dissipative quantum many-body dynamics [see Eqs.~\eqref{recurrence},\eqref{lindbladian}], some robustness to noise can be expected. 
While our simulations are ultimately limited by bottlenecks of classical devices, our findings provide evidence for the practical viability of large-scale QNNs and offer a starting point for their  systematic implementation on real NISQ devices.

\begin{acknowledgements}
\noindent \textbf{Acknowledgments.} 
We thank F. Carollo and G. Perfetto for discussions. The research leading to these results has received funding from the Deutsche Forschungsgemeinschaft (DFG, German Research Foundation) under Project No. 449905436, as well as through the Research Unit FOR 5413/1, Grant No. 465199066. This Project has also received funding from the European Union’s Horizon Europe research and innovation program under Grant Agreement No. 101046968 (BRISQ), and from EPSRC under Grant No. EP/V031201/1. We acknowledge the financial support by the German Federal Ministry of Research, Technology and Space (BMFTR) within the project "Neuronale Quantennetzwerke auf NISQ-Quantencomputern (NeuQuant)" under grant 13N17065. This work is supported by the ERC grant OPEN-2QS (Grant No.\ 101164443, https://doi.org/10.3030/101164443). IL is a member of the Machine Learning Cluster of Excellence, funded by the Deutsche Forschungsgemeinschaft (DFG, German Research Foundation) under Germany’s Excellence Strategy—EXC Number 2064/1 - Project Number 390727645. The authors acknowledge support by the state of Baden-Württemberg through bwHPC and the German Research Foundation (DFG) through grant no INST 40/575-1 FUGG (JUSTUS 2 cluster).
\end{acknowledgements}

\bibliography{Reference}

\setcounter{equation}{0}
\setcounter{figure}{0}
\setcounter{table}{0}
\renewcommand{\theequation}{S\arabic{equation}}
\renewcommand{\thefigure}{S\arabic{figure}}

\makeatletter
\renewcommand{\theequation}{S\arabic{figure}}
\renewcommand{\thefigure}{S\arabic{figure}}

\onecolumngrid
\newpage

\setcounter{page}{1}

\setcounter{secnumdepth}{3}
\pagestyle{plain}

\begin{center}
{\Large SUPPLEMENTAL MATERIAL}
\setcounter{page}{1}
\end{center}
\begin{center}
\vspace{0.8cm}
{\Large Getting large-scale quantum neural networks ready for quantum hardware}
\end{center}
\begin{center}
Mario Boneberg,$^{1}$, Simon Kochsiek$^{1}$ and Igor Lesanovsky$^{1,2}$
\end{center}
\begin{center}
$^1$ {\em Institut f\"ur Theoretische Physik, Universit\"at Tübingen and Center for Integrated Quantum Science and Technology,}\\
{\em  Auf der Morgenstelle 14, 72076 T\"ubingen, Germany}\\
$^2$ {\em School of Physics and Astronomy and Centre for the Mathematics}\\
{\em  and Theoretical Physics of Quantum Non-Equilibrium Systems,}\\
{\em  The University of Nottingham, Nottingham, NG7 2RD, United Kingdom}
\end{center}

\setcounter{equation}{0}
\setcounter{figure}{0}
\setcounter{table}{0}
\setcounter{page}{1}
\makeatletter
\renewcommand{\theequation}{S\arabic{equation}}
\renewcommand{\thefigure}{S\arabic{figure}}

\makeatletter
\renewcommand{\theequation}{S\arabic{equation}}
\renewcommand{\thefigure}{S\arabic{figure}}

\renewcommand{\bibnumfmt}[1]{[S#1]}
\renewcommand{\citenumfont}[1]{S#1}

\onecolumngrid

\setcounter{secnumdepth}{3}


\section{Parameter update rule: nadam stochastic gradient descent} \label{sec1}
In this section, we explain in detail how the parameter matrices $\mathbf{h}$ and $\mathbf{j}$, defined in Eqs.~(5)-(6) of the main text, are updated in our learning algorithm. As already presented in the main text, this succeeds via a variant of stochastic gradient descent, where we randomly pick a subset of $P'$ states of the dataset $\{ \rho_{\nu_i}^{(i)} \}_{i=1}^{P}$ and compute the negative gradient of the loss $\Vec{\nabla}L$ via Eq.~(9) of the main text. We then move by a small amount --- controlled by the learning rate $\lambda$ and the function $f_r$ --- into the direction of fastest decrease with respect to this subset via [see Eq.~(10) of the main text]
\begin{equation} \label{SM0}
(\mathbf{h}_r, \mathbf{j}_r) \longrightarrow (\mathbf{h}_r, \mathbf{j}_r) - \lambda f_r(\nabla L(\mathbf{h}_r, \mathbf{j}_r)) .
\end{equation}
Here we indicated the current training round $r$ in the parameters for which the function $f_r$ is defined recursively \cite{ruder2016}. It weights the gradient with respect to the current $P'$ and parametrization with past gradients according to
\begin{equation} \label{SM1}
    m_r = \beta_1 m_{r-1} + (1 - \beta_1) \nabla L(\mathbf{h}_r, \mathbf{j}_r)) 
\end{equation}
in order to keep some memory about the gradient direction. In addition, it tracks the sizes of the gradients via 
\begin{equation} \label{SM2}
    v_r = \beta_2 v_{r-1} + (1 - \beta_2) (\nabla L(\mathbf{h}_r, \mathbf{j}_r))^2 ,
\end{equation}
where the square has to be understood elementwise.
Initially we set $m_0=0, v_0=0$. This results in $m_r,v_r$ which are biased towards zero, especially early in training. This is accounted for by introducing the bias-correction factors $(1- \beta_1^r)^{-1}$ and $(1- \beta_2^r)^{-1}$ which yield
\begin{equation}
    \hat{m}_r = \frac{m_r}{1- \beta_1^r}
\end{equation}
and
\begin{equation}
    \hat{v}_r = \frac{v_r}{1- \beta_2^r}.
\end{equation}
Putting this together, parameters are updated as 
\begin{equation}
(\mathbf{h}_r, \mathbf{j}_r) \longrightarrow (\mathbf{h}_r, \mathbf{j}_r) - \lambda \frac{\hat{m}_r}{\sqrt{\hat{v}_r } + \delta} .
\end{equation}
In this formula, division has to be understood elementwise. Average gradient directions accumulate and are rescaled by their size. Rescaling imposes an adaptive learning rate on the individual parameters. The small number $\delta$ is introduced for numerical stability. As a last step, we incorporate a lookahead in the update, i.e. we compute the update based on approximating the future direction of the gradient by replacing $m_{r-1} \rightarrow m_r$ in Eq.~\eqref{SM1} which yields Eq.~\eqref{SM0} with 
\begin{equation}
    f_r(\nabla L(\mathbf{h}_r, \mathbf{j}_r)) = \frac{ \beta_1 \hat{m}_r + \frac{1-\beta_1}{1-\beta_1^r} \nabla L(\mathbf{h}_r, \mathbf{j}_r) }{\sqrt{\hat{v}_r } + \delta}.
\end{equation}
Further below, we discuss which values we fixed for the parameters $\beta_1, \beta_2,\lambda, \delta$ in order to arrive at the results presented in the main text.

\section{Data sampling}
Here we want to show how we sample both of the datasets used in the main text. They correspond to translation invariant pure product states
\begin{equation}
    \rho = \bigotimes_{k=1}^N \outerproduct{\psi}{\psi} , \qquad 
\end{equation}
that can be labelled by either A or B. We first randomly select such a label with equal probabilities. For the magnetization observables $\hat{m}^\alpha =\sum_{k=1}^N \sigma_k^\alpha / (2N)$ and their expectation values $m^\alpha = \Tr (\hat{m}^\alpha \rho )$, these pure, translation invariant product states have to fulfill the equation
\begin{equation}\label{SM3}
    \left( m^x \right)^2 + \left( m^y \right)^2 + \left( m^z \right)^2 = \frac{1}{4} .
\end{equation}
That is, the states lie on the Bloch sphere.
\subsection{Dataset I}
For the first dataset, for label A, we draw a value $m^z$ from a uniform distribution in the interval $[0.15,0.4]$. We then choose the other components on the Bloch sphere by computing $R=\sqrt{1 /4 - \left( m^z \right)^2  }$ and defining
\begin{equation}
     m^x = R \cos \alpha,\qquad m^y = R \sin \alpha
\end{equation}
with $\alpha$ chosen uniformly in the interval $[0,2 \pi ]$ such that Eq.~\eqref{SM3} is fulfilled. The state $\rho$ can be parametrized via 
\begin{equation}
    \ket{\psi} = \cos(\frac{\theta}{2}) |0 \rangle + e^{i \phi} \sin(\frac{\theta}{2}) |1 \rangle 
\end{equation}
with the angles $\theta \in [0,\pi]$ and $\phi \in [0, 2 \pi ]$. We determine this angles through the magnetizations $m^\alpha$ via the relations 
\begin{equation}
    \theta = \arccos(2m^z) \ , \quad \phi = \mathrm{arctan2}(m^y,m^x) 
\end{equation}
with the function
\begin{equation}
\mathrm{arctan2}(y,x) =
\begin{cases}
\arctan\!\left(\tfrac{y}{x}\right), & x > 0, \\[1em]
\arctan\!\left(\tfrac{y}{x}\right) + \pi, & x < 0,\; y > 0, \\[1em]
\pm \pi, & x < 0,\; y = 0, \\[1em]
\arctan\!\left(\tfrac{y}{x}\right) - \pi, & x < 0,\; y < 0, \\[1em]
+\tfrac{\pi}{2}, & x = 0,\; y > 0, \\[0.5em]
-\tfrac{\pi}{2}, & x = 0,\; y < 0,
\end{cases}
\end{equation}
which has to be used as the underlying expression for $\phi \in (-\pi,\pi]$ is $\tan(\phi)=m^y / m^x$. The inverse of $\tan$ does not distinguish the cases with, e.g., $\mathrm{sign}(m^y)=1, \mathrm{sign}(m^x)=1$ and $\mathrm{sign}(m^y)=-1, \mathrm{sign}(m^x)=-1$. The $\mathrm{arctan2}$ accounts for this and gives the correct quadrant for the angle $\phi$. For label B, on the other hand, we draw a value $m^z$ from a uniform distribution in the interval $[-0.4, -0.15]$ and build the states in the same way as explained above.
\subsection{Dataset II}
For the second dataset, for both labels, we randomly selected two values for $m^z$, corresponding to $\theta =\pi / 4$ and $\theta =\frac{3 \pi}{4}$, with equal probabilities. For label A, if $\theta =\pi / 4$ we sample $\phi$ in the interval 
\begin{equation}
    \mathcal{I}_1 = \left[ \frac{ \pi}{4} , \frac{3 \pi}{4} \right]
\end{equation}
and if $\theta =\frac{3 \pi}{4}$ we sample it in the interval 
\begin{equation}
    \mathcal{I}_2 = \left[ \frac{5 \pi}{4} , \frac{7 \pi}{4} \right].
\end{equation}
For label B, we sample $\phi$ in $\mathcal{I}_1$ if $\theta =\frac{3 \pi}{4}$ and in $\mathcal{I}_2$ if $\theta =\pi / 4$.

\section{Training parameters and loss function behavior}
\begin{figure*}[t]
    \centering
    \includegraphics[width=\textwidth]{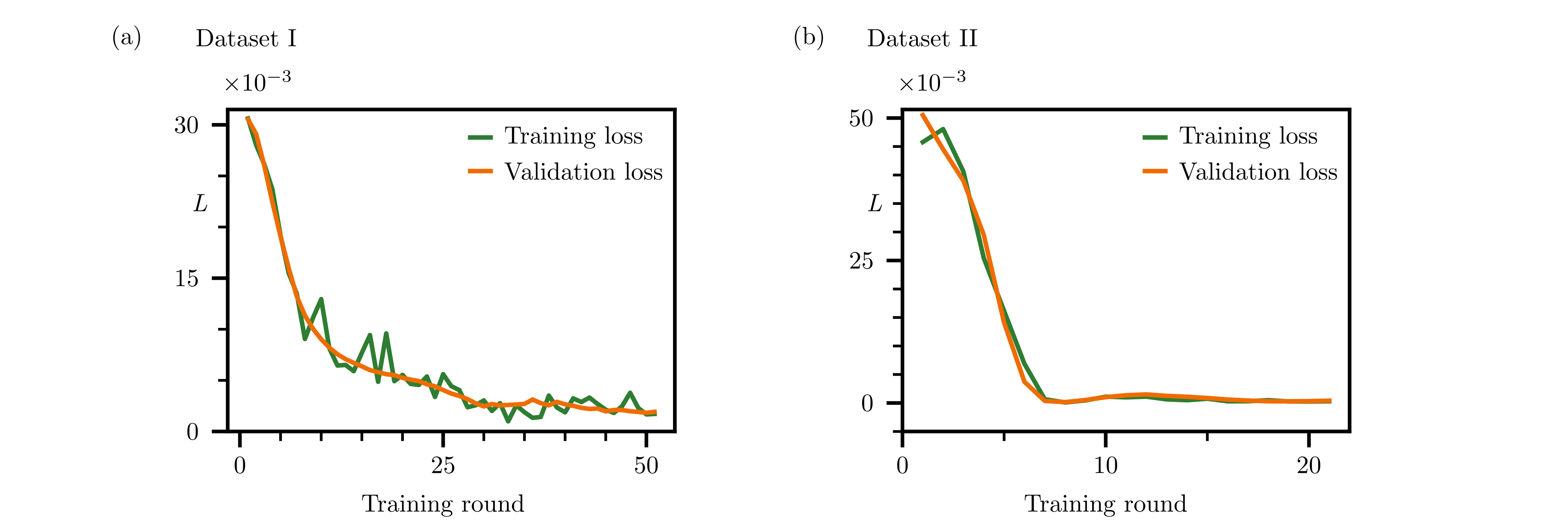} 
\caption{\textbf{Loss behavior during training.} (a) Training loss (green) and validation loss (orange) for dataset I. Both losses decrease, on average, over the $50$ training rounds with the validation loss being minimal after training round $49$. As training is carried out on different minibatches, the training loss exhibits oscillations. (b) Training loss and validation loss for dataset II. Both losses decrease fast. The validation loss is minimal after training round $7$ of the $20$ training rounds.}
    \label{SM:fig1}
\end{figure*}

In this section, we want to provide additional information on the training behavior, parameters and hyperparameters used to produce the simulation results presented in Fig.~2 of the main text. In Section \ref{sec1}, we explained the stochastic gradient variant, which involves the hyperparameters $\beta_1, \beta_2, \lambda, \delta$ and also the minibatch size $P'$. For the first dataset, shown in Fig.~2(a) of the main text, we have fixed 
\begin{equation}
    \beta_1 = 0.75  , \quad \beta_2 = 0.98 , \quad \lambda = 0.05 , \quad \delta = 10^{-7} 
\end{equation}
and a parameterization of $H_{k,\ell - 1}$ and $J_{k,\ell - 1}$ [see Eqs.(5),(6) of the main text]  
\begin{equation}
h=
    \begin{pmatrix}
        0 & a_1 & 0 & a_2 \\
        0 & a_3 & 0 & 0 \\
        0 & 0 & a_4 & 0 \\
        0 & 0 & 0 & a_5 
    \end{pmatrix}
    \qquad
    \mathrm{and}
    \qquad
    j=
    \begin{pmatrix}
        0 & a_6 & a_7 + i a_8 & a_9 + i a_{10} \\
        0 & 0 & 0 & 0 \\
        0 & 0 & 0 & 0 \\
        0 & 0 & 0 & 0 
    \end{pmatrix},\label{eq:params1}
\end{equation}
with the 10-component real parameter vector $\Vec{a}$, which we initially fixed as
\begin{equation}
    \Vec{a}_{\mathrm{initial}} = (1, -1, 1, 1, -1, 1, -1, -1, 1, -1)^T .
\end{equation}
We randomly chose $P=250$ training data states and trained the network on randomly chosen minibatches with $P'=25$ states. In Fig.~\ref{SM:fig1}(a), we show the loss on these minibatches over the training rounds. Moreover, we plot the validation loss during training, i.e. the loss computed on $50$ data states that are not seen during training. As can be seen, both losses decrease and for the round with the smallest validation loss, we verified in Fig.~2(c) in the main text that the corresponding QNN distinguishes the classes on the output layer. This network is parameterized as 
\begin{equation}
    \Vec{a}_{\mathrm{trained}} \approx (0.32, 0.9, -0.38, -0.26, -1.08, 0.27, -0.43, -0.17, 0.38, -0.64)^T .
\end{equation}
For the second dataset, shown in Fig.~2(d) of the main text, we have fixed 
\begin{equation}
    \beta_1 = 0.85  , \quad \beta_2 = 0.9995 , \quad \lambda = 0.05 , \quad \delta = 10^{-7} 
\end{equation}
and a parameterization 
\begin{equation}
h=
    \begin{pmatrix}
        0 & b_1 & 0 & 0 \\
        0 & b_2 & 0 & 0 \\
        0 & 0 & b_3 & 0 \\
        0 & 0 & 0 & b_4 
    \end{pmatrix}
    \qquad
    \mathrm{and}
    \qquad
    j=
    \begin{pmatrix}
        0 & b_5 & 0 & b_6 \\
        0 & b_7 & b_8 & b_9 \\
        0 & b_{10} & 0 & 0 \\
        0 & b_{11} & 0 & 0 
    \end{pmatrix}.\label{eq:params2}
\end{equation}
The 11-component real parameter vector $\Vec{b}$ we fixed initially as 
\begin{equation}
    \Vec{b}_{\mathrm{initial}} \approx (0, -1, -1, -1  , 0, 0, 0, -1, 0, 0, 0)^T .
\end{equation}
Also for this dataset we choose $P=250$ training states and $50$ validation states, but minibatches of size $P'=23$. The losses decrease, as illustrated in Fig.~\ref{SM:fig1}(b) and at the minimal validation loss the QNN is parameterized as [cf. main text Fig.~2(f)]
\begin{equation}
    \Vec{b}_{\mathrm{trained}} \approx (-0.05, -1.06, -0.55, -1.33, 0.01, -0.40, -0.07, -0.55, -0.04, 0.26, -0.03)^T .
\end{equation}

\section{Details on Tensor Networks}
Here we want to explain how we use tensor networks for simulation of our QNNs in order to arrive at the results presented in Fig.~2 of the main text.
\subsection{Tensor network simulation techniques}
As we deal with quantum systems which have $550$ qubits, states live in a Hilbert space which has dimension $\approx 10^{331}$. To simulate the evolution of such systems, it is therefore necessary to reduce the number of degrees of freedom and we do this by employing tensor network descriptions where we discard irrelevant information by singular value truncations. In the following we want to briefly discuss how this works; for a fully detailed explanation see Refs.~\cite{boneberg2025,gillman2021b}. 

The output state of the QNN is given by the relation [see also main text Eq.~(1)]
\begin{equation}
    \rho_{\mathrm{out}} = \Tr_{\neq L} \left( \prod_{\ell=L}^1 \prod_{k=1}^N  G_{k,\ell}  \rho_{\mathrm{in}} \otimes \outerproduct{\mathbf{0}}_{\neq 0} \prod_{\ell=1}^L \prod_{k=N}^1  G_{k,\ell}^\dagger  \right)
\end{equation}
with the local gates $G_{k,\ell}$, defined in the main text. As these only act on a neighborhood of site $k$ in layers $\ell - 1$ and $\ell$, we may equivalently express this input-output relation via the recurrence relation 
\begin{equation}\label{SM:recurrence}
    \rho_{\ell} = \Tr_{\ell - 1} \left( \prod_{k=1}^N  G_{k,\ell} \rho_{\ell - 1} \otimes \outerproduct{\mathbf{0}}{\mathbf{0}}_\ell  \prod_{k=N}^1  G_{k,\ell}^\dagger  \right) ,
\end{equation}
where $\Tr_{\ell -1}$ denotes the trace over layer $\ell -1$, $\rho_\ell, \rho_{\ell -1}$ the states on layers $\ell, \ell -1$, and $G_{k, \ell}$ the operators which are defined only on layers $\ell -1, \ell$. The state $\outerproduct{\mathbf{0}}{\mathbf{0}}_\ell $ is the vacuum state on layer $\ell$. Input states $\rho_{\mathrm{in}}= \rho_0$ are propagated to the output state $\rho_{\mathrm{out}}= \rho_L$. Equation \eqref{SM:recurrence} greatly reduces the amount of resources needed to simulate the network evolution as states now live only in a Hilbert space which has dimension $16^N$. However, for our systems with $N=50$ qubits in each of the $L+1=11$ layers, this corresponds to a dimension $\approx 10^{60}$, which is still large. 

In order to apply our tensor network algorithms, we vectorize Eq.~\eqref{SM:recurrence} which gives us 
\begin{equation} \label{SM:recurrencevec}
    \ket{\rho_\ell } = F^\ell \ket{\rho_{\ell - 1}}
\end{equation}
with a $4^N \times 4^N$-dimensional operator $F^\ell$ which acts on the $4^N$-dimensional ket state $\ket{\rho_{\ell -1}}$. Our input states are such that they can be represented as matrix-product states (MPSs)
\begin{equation}
    \ket{\rho_{\mathrm{in}}} = \sum_{ \sigma_{i, \ell -1} , \sigma_{i, \ell } = 0,1} M^{[1], \sigma_{1, \ell -1} \sigma_{1, \ell }} M^{[2], \sigma_{2, \ell -1} \sigma_{2, \ell }} \cdots M^{[N], \sigma_{N, \ell -1} \sigma_{N, \ell }} \ket{\sigma_{1, \ell -1} \sigma_{1, \ell } \sigma_{2, \ell -1} \sigma_{2, \ell } \ldots \sigma_{N, \ell -1} \sigma_{N, \ell }}
\end{equation} 
with small (bond) dimensions of the matrices $M^{[i], \sigma_{i, \ell -1} \sigma_{i, \ell }}$ and thus fit into a classical computer. In addition, we can built the operator $F^\ell$ based on the local gates as a matrix-product operator (MPO) \cite{boneberg2025} 
\begin{align}
    F^\ell =& \sum_{ \sigma_{i, \ell -1} , \sigma_{i, \ell }, \sigma'_{i, \ell -1} , \sigma'_{i, \ell } = 0,1} O^{[1], \sigma'_{1, \ell -1}  \sigma'_{1, \ell } \sigma_{1, \ell -1} \sigma_{1, \ell }} O^{[2], \sigma'_{2, \ell -1}  \sigma'_{2, \ell } \sigma_{2, \ell -1} \sigma_{2, \ell }} \cdots O^{[N], \sigma'_{N, \ell -1}  \sigma'_{N, \ell } \sigma_{N, \ell -1} \sigma_{N, \ell }} \\
    & \times \ket{\sigma'_{1, \ell -1} \sigma'_{1, \ell } \sigma'_{2, \ell -1} \sigma'_{2, \ell } \ldots \sigma'_{N, \ell -1} \sigma'_{N, \ell }} \bra{\sigma_{1, \ell -1} \sigma_{1, \ell } \sigma_{2, \ell -1} \sigma_{2, \ell } \ldots \sigma_{N, \ell -1} \sigma_{N, \ell }} .
\end{align} 
We thus simulate the QNN evolution by repeatedly applying the MPO to the input MPS in order to get the output MPS $\ket{\rho_{\mathrm{out}}}$. As the maximum matrix (bond) dimension $\chi$, which quantifies the resources needed to represent the state, increases with each application of the MPO, we use standard SVD truncation to discard irrelevant degrees of freedom and keep $\chi$ small. 

\begin{figure*}[b]
    \centering
    \includegraphics[width=\textwidth]{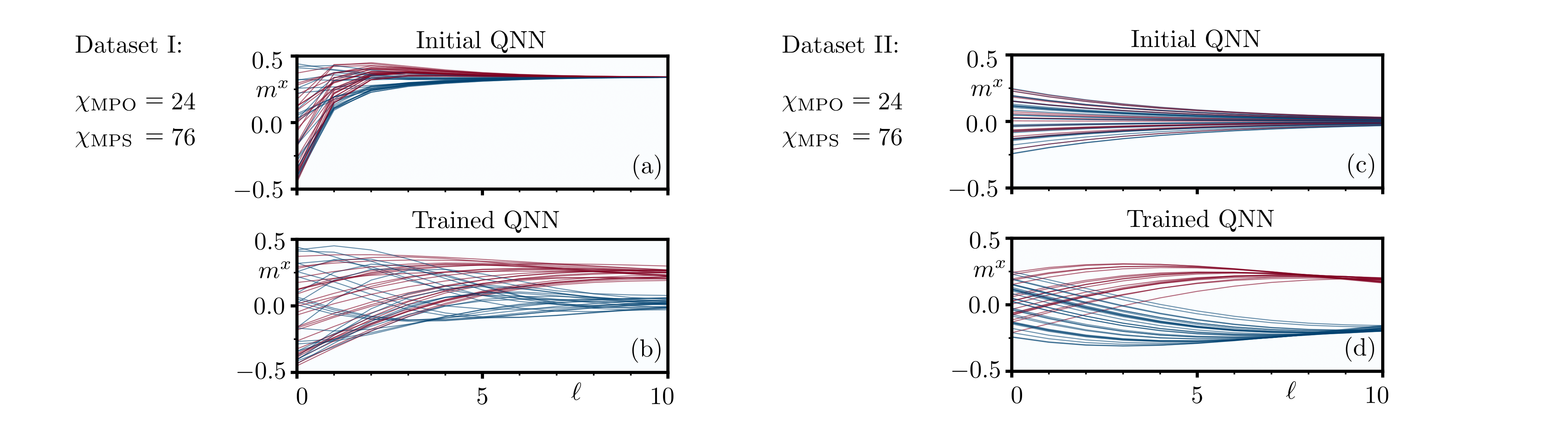} 
\caption{\textbf{Tensor network checks of QNN evolution.} (a) Evolution of $m^x$ through the layers of the initial QNN for validation states of dataset I. (b) Evolution of the trained QNN for dataset I. (c) Initial QNN evolution for validation states of dataset II. (d) Trained QNN evolution for dataset II. We simulated the QNNs using bond dimensions $\chi_{\mathrm{MPO}}=24$ and $\chi_{\mathrm{MPS}}=76$. The behavior of the trajectories agrees with the one for the smaller bond dimensions shown in Fig.~2 of the main text. Trained networks classify the states according to their labels (red and blue, respectively).}
    \label{SM:fig2}
\end{figure*} 

From the output MPSs $\ket{\rho^{(i)}_{\mathrm{out}}}$ for all input data states $i=1,\ldots, P'$, we want to compute the loss function [see main text Eq.~(8)] for $5000$-shot measurements of the magnetization observable $\hat{m}^x= \sum_{k=1}^N \sigma_k^x / (2N)$. A single-shot thus corresponds to projective measurements $P_1 = \outerproduct{+}{+}$, $P_2 = \outerproduct{-}{-}$ on every site $k$, where $\ket{+},\ket{-}$ are the eigenstates of $\sigma^x$ with the eigenvalues $\pm 1$; the measurement outcomes are then averaged over all sites. In our simulations we do this as follows. We first compute the probability of measuring outcome $m_1=\pm 1$ on the first qubit (we implicitly sum over repeated indices and for ease of notation omit the superscript $[i]$ in tensors)
\begin{align}
    p_1(m_1) =& \Tr (P_1^{(m_1)} \rho) = \sum_{\Vec{\epsilon}, \Vec{\sigma}(')} M^{\sigma_1' \sigma_1} M^{\sigma_2' \sigma_2} \cdots M^{\sigma_N' \sigma_N}  \bra{\Vec{\epsilon}} \outerproduct{m_1}{m_1} \otimes \mathbb{I} \ket{\Vec{\sigma}'} \bra{\Vec{\sigma}} \ket{\Vec{\epsilon}} \\ 
    =& T^{m_1} M^{\epsilon_2 \epsilon_2} \cdots M^{\epsilon_N \epsilon_N}
\end{align}
and sample a result $m_1$ from this distribution. We then compute the conditional probability of measuring outcome $m_2=\pm 1$ on the second qubit 
\begin{align}
    p_2(m_2 | m_1) =& \frac{\Tr (P_1^{(m_1)} P_2^{(m_2)} \rho)}{p_1(m_1)}  = T^{m_1 } T^{m_2} M^{\epsilon_3 \epsilon_3} \cdots M^{\epsilon_N \epsilon_N} / T^{m_1} M^{\epsilon_2 \epsilon_2} \cdots M^{\epsilon_N \epsilon_N}
\end{align}
and sample a result $m_2$. We iterate this up to the $N$-th site where we get 
\begin{align}
    p_N(m_N | m_1,m_2,\ldots,m_{N-1}) = T^{m_1 } T^{m_2} \cdots T^{m_N} / T^{m_1} T^{m_2} T^{m_{N-1}}  \cdots M^{\epsilon_N \epsilon_N}
\end{align}
and sample $m_N$. Here we denoted 
\begin{equation}
    T^{m_i} = \begin{cases}
        \frac{1}{2} \left( M^{[i],00} + M^{[i],01} + M^{[i],10} + M^{[i],11} \right) & m_i=+1 \\
        \frac{1}{2} \left( M^{[i],00} - M^{[i],01} - M^{[i],10} + M^{[i],11} \right) & m_i=-1   .
    \end{cases}
\end{equation}
Sampling the outcomes $m_k$ in this way corresponds to sampling them from the full probability distribution $p(m_1,m_2,\ldots,m_N)=\Tr (P_1^{(m_1)} P_2^{(m_2)} \cdots P_N^{(m_N)} \rho)$ as 
\begin{equation}
    p(m_1,m_2,\ldots,m_N) = p_1(m_1) p_2(m_2 | m_1) p_3( m_3 | m_2,m_1) \cdots p_N ( m_N | m_{N-1}, m_{N-2} , \ldots , m_1 ) .
\end{equation}
We compute the magnetization $m^x$ by averaging over all shots and all sites 
\begin{equation}
    m_{S,\mathrm{out}}^{x} = \frac{1}{2SN} \sum_{s=1}^S \sum_{k=1}^N m_k^s ,
\end{equation}
where $m_k^s$ denotes the outcome of shot $s$ on site $k$.

\subsection{Convergence of simulations}
\begin{figure*}[b]
    \centering
    \includegraphics[width=\textwidth]{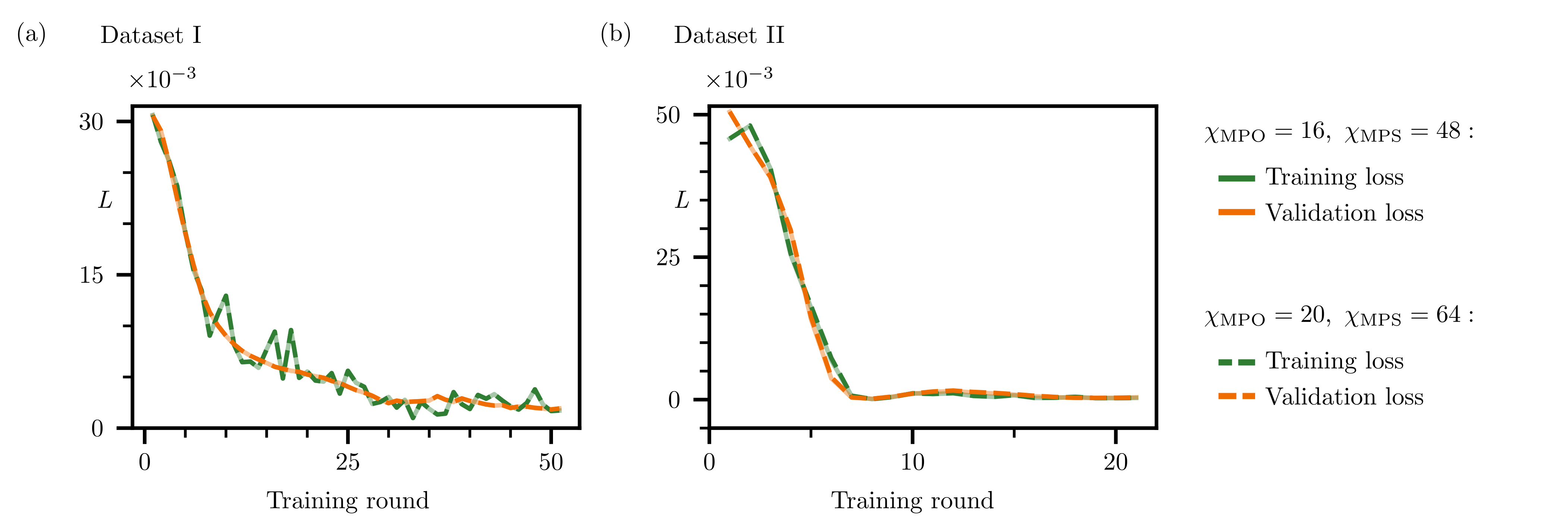} 
\caption{\textbf{Tensor network checks for loss functions.} (a) Training loss (green) and validation loss (orange) during training for dataset I. (b) Losses for dataset II. The loss functions for bond dimensions $\chi_{\mathrm{MPO}}=20$ and $\chi_{\mathrm{MPS}}=64$ (dashed lines) closely align with the loss functions for $\chi_{\mathrm{MPO}}=16$ and $\chi_{\mathrm{MPS}}=48$ (full lines), indicating that the truncation error is negligible.}
    \label{SM:fig3}
\end{figure*} 
We train our networks on Datasets I and II by evolving minibatches of size $P'=25$ and $P'=23$, respectively, for the randomly sampled state full batch $\{\rho_{\nu_i}^{(i)}\}_{i=1}^{P=250}$ with $\nu_i \in \{ \mathrm{A, B} \}$. From the output states we compute the gradient of the loss function [see Eqs.~(8),(9) of the main text] 
\begin{equation}
    \nabla_{\alpha_1, \alpha_2, \beta} L ( \mathbf{h}, \mathbf{j} ) \approx  \frac{ L( (\mathbf{h}, \mathbf{j} ) + \epsilon \mathbf{E}^\beta_{\alpha_1 \alpha_2} ) - L( \mathbf{h}, \mathbf{j} )}{\epsilon} 
\end{equation}
with the finite-difference parameter $\epsilon $, $\beta =0 \ \mathrm{or} \ 1$, $\mathbf{E}_{\alpha_1 \alpha_2}^1 = (\mathbf{E}_{\alpha_1 \alpha_2},0)$, $\mathbf{E}_{\alpha_1 \alpha_2}^2 = (0, \mathbf{E}_{\alpha_1 \alpha_2})$ and the matrix $\mathbf{E}_{\alpha_1 \alpha_2}$ with a one at position $(\alpha_1, \alpha_2)$ and zeros everywhere else. We update the parameter matrices $\mathbf{h}, \mathbf{j}$ in the optimal direction [see Eq.~(10) of the main text and Sec.~\ref{sec1}]
\begin{equation}
(\mathbf{h}, \mathbf{j}) \longrightarrow (\mathbf{h}, \mathbf{j}) - \lambda f_r(\nabla L(\mathbf{h}, \mathbf{j}))
\end{equation}
by means of the learning rate $\lambda$ and the function $f_r$ depending on the current training round $r$. After each round we assess the classification capability of the network by evolving a fixed set of $50$ randomly sampled validation states, labeled by A and B.

\begin{figure*}[t]
    \centering
    \includegraphics[width=\textwidth]{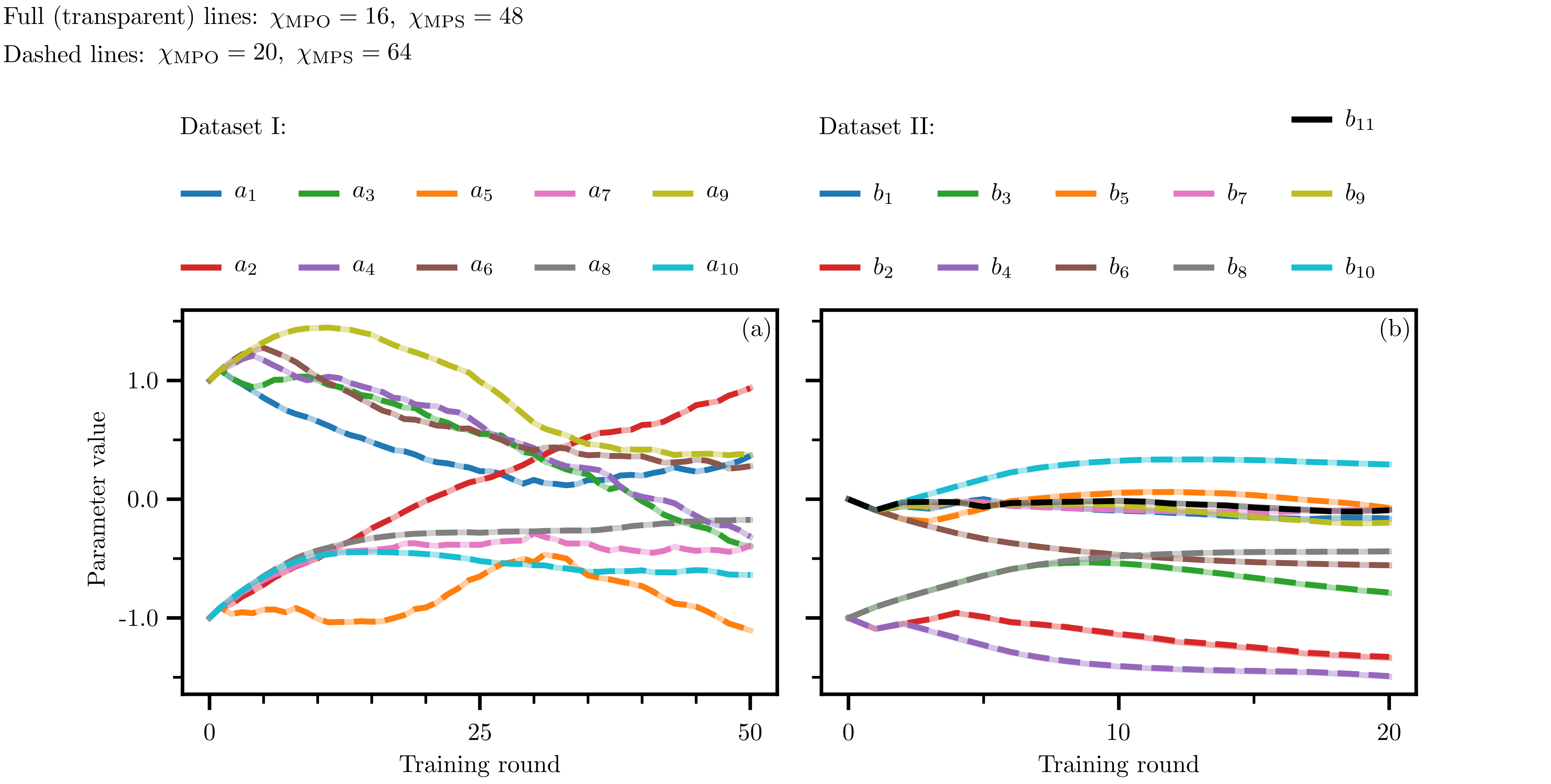} 
\caption{\textbf{Tensor network checks on the training.} (a) Training parameters during training for dataset I. (b) Training parameters for dataset II. The training trajectories for bond dimensions $\chi_{\mathrm{MPO}}=20$ and $\chi_{\mathrm{MPS}}=64$ (dashed lines) overlap with the ones for $\chi_{\mathrm{MPO}}=16$ and $\chi_{\mathrm{MPS}}=48$ (full lines), illustrating that the gradient and update the algorithm computes remains meaningful when truncating the bond dimension.
The components $a_i$ and $b_i$ are the entries of the parameter matrices $\mathbf{h},\mathbf{j}$ for datasets I and II \eqref{eq:params1} and \eqref{eq:params2}.}
    \label{SM:fig4}
\end{figure*}

We evolve the states by employing Eq.~\eqref{SM:recurrencevec} and get the results shown in Fig.~2 of the main text. For this we have truncated the operator $F^\ell$ via SVD compression to have maximum matrix bond dimension $\chi_{\mathrm{MPO}}=16$ and used an MPS bond dimension $\chi_{\mathrm{MPS}}=48$. Here we want to demonstrate that truncation to those values for the bond dimensions does not qualitatively alter the results and captures the main physics. Indeed, in Fig.~\ref{SM:fig2}(a-d) we plot the evolutions of the validation states for both datasets and for initial and trained QNN for higher bond dimensions, $\chi_{\mathrm{MPO}}=24$ and $\chi_{\mathrm{MPS}}=76$. A comparison with Fig.~2 of the main text shows that these results look almost identical to the results we obtained for smaller bond dimension. This is already strong evidence that the errors from singular value truncations are sufficiently small such that they do not qualitatively change our simulations. The trained network remains classifying when increasing accuracy, i.e. the bond dimension. To show convergence with bond dimension, however, we also need to demonstrate that the output the QNN computes during training remains invariant when increasing the bond dimension. For this purpose, we trained the network again by employing throughout bond dimensions $\chi_{\mathrm{MPO}}=20$ and $\chi_{\mathrm{MPS}}=64$. We display in Fig.~\ref{SM:fig3}(a-b) the validation and training losses during training for both datasets. Their behaviors agree with the ones for bond dimensions $\chi_{\mathrm{MPO}}=16$ and $\chi_{\mathrm{MPS}}=48$, shown in Fig.~\ref{SM:fig1}. Moreover, in Fig.~\ref{SM:fig4}(a-b) we also plot the parameters during training for both datasets. The curves align closely for both bond dimension settings, indicating that also the gradients and the updates are computed faithfully.

This shows that the results of our training are not changed qualitatively when employing larger bond dimensions. There may, however, still be small numeric inaccuracies. These numerical deviations do not diminish the relevance of our results. Indeed, in the computation of the gradient
\begin{equation}
    \nabla_{\alpha_1, \alpha_2, \beta} L ( \mathbf{h}, \mathbf{j} ) \approx  \frac{ \tilde{L}( (\mathbf{h}, \mathbf{j} ) + \epsilon \mathbf{E}^\beta_{\alpha_1 \alpha_2} ) - \tilde{L}( \mathbf{h}, \mathbf{j} )}{\epsilon} ,
\end{equation}
the finite-bond dimension estimates $\tilde{L}( (\mathbf{h}, \mathbf{j} ) + \epsilon \mathbf{E}^\beta_{\alpha_1 \alpha_2} )$ and $\tilde{L}( \mathbf{h}, \mathbf{j} )$ can be interpreted as 
\begin{align*}
    \tilde{L}( (\mathbf{h}, \mathbf{j} ) + \epsilon \mathbf{E}^\beta_{\alpha_1 \alpha_2} ) &= L( (\mathbf{h}, \mathbf{j} ) + \epsilon \mathbf{E}^\beta_{\alpha_1 \alpha_2} ) + \eta_1 , \\
    \tilde{L}( \mathbf{h}, \mathbf{j} ) &= L( \mathbf{h}, \mathbf{j} ) + \eta_2 
\end{align*}
That is, as an original signal without singular value truncations plus \textit{systematic} inaccuracies $\eta_i$, originating from singular value truncations. But this is exactly how inaccuracies due to noise would also enter when estimating the gradient on a real NISQ device. Therefore, our results can be seen as providing evidence for trainability under the impact of systematic inaccuracies. In the end the trained network classifies initial states successfully.

\end{document}